\documentclass[12pt]{article}

\usepackage[margin=1in]{geometry}
\usepackage[T1]{fontenc}
\usepackage[utf8]{inputenc}
\usepackage{lmodern}
\usepackage[american]{babel}
\usepackage{csquotes}
\usepackage{hyperref}
\usepackage{graphicx}
\usepackage{subcaption}
\usepackage{wrapfig}

\usepackage[
backend=biber,
style=apa,
sorting=nyt
]{biblatex}

\title{HARP: The Human--AI Research Platform}

\author{
    Zeshu Zhu\\
    BTPX Innovation Lab
    \and
    Natalie Friedman\\
    BTPX Innovation Lab
    \and
    Kevin Weatherwax\\
    BTPX Innovation Lab
    \and
    Emily Eiben\\
    BTPX User Assistance
}

\date{}

\begin{document}

\maketitle

\begin{center}
Presented at the SAP Academic Community Conference (SAP ACC), July 2026
\end{center}
\vspace{2.5em}

\section{Abstract}

Large language models (LLMs) have shifted human--computer interaction from `traditional'' interface journeys toward more conversational exchanges.
Researchers studying HCI and UI use moderated usability sessions, interviews, surveys, transcript analysis, and static prototypes.
However, static prototypes provide limited opportunities to study interaction with live AI systems or systematically control how an LLM behaves across participants and scenarios.
Conversation transcripts reveal little about how users formulate, revise, and hesitate over prompts before submission.
We designed the Human--AI Research Platform (HARP) for researchers, designers, and anyone who has ever wondered, `What if AI did this?'
HARP places participants in controlled mock scenarios with live, configurable AI agents.
Researchers can control agent prompts, model parameters, response characteristics, and experimental conditions; trigger surveys at predefined moments; and record prompt composition time, response latency, deletions, and keystroke pauses.
Planned capabilities include voice, facial expression, gesture, and, where legally and ethically appropriate, emotion analysis.
We illustrate HARP through a study examining how technical specificity and response length affect retention of LLM output.
By pairing controllable live agents with behavioral and self-report measures, HARP enables systematic testing of how AI design choices affect users.

\section{Introduction}

Large language models (LLMs) have created a new paradigm for human--computer interaction.
Researchers often study these interactions through moderated usability sessions, interviews, surveys, and conversation transcript analysis.
However, studying human interaction with live LLMs introduces a methodological challenge: researchers need participants to engage with responsive AI systems while maintaining control over how those systems behave across participants and experimental conditions.
Static prototypes and screenshots can provide this control, but they cannot reproduce the dynamic and open-ended nature of interaction with a live LLM, potentially reducing ecological validity.
Researchers who use live systems may instead have limited control over the responses participants encounter, making it difficult to isolate the effects of particular prompts, response characteristics, or agent behaviors.
Moreover, conversation logs alone provide limited visibility into how users formulate prompts, revise requests, and hesitate while engaging with AI-generated responses.

\begin{wrapfigure}{r}{0.48\columnwidth}
\centering
\includegraphics[width=\linewidth]{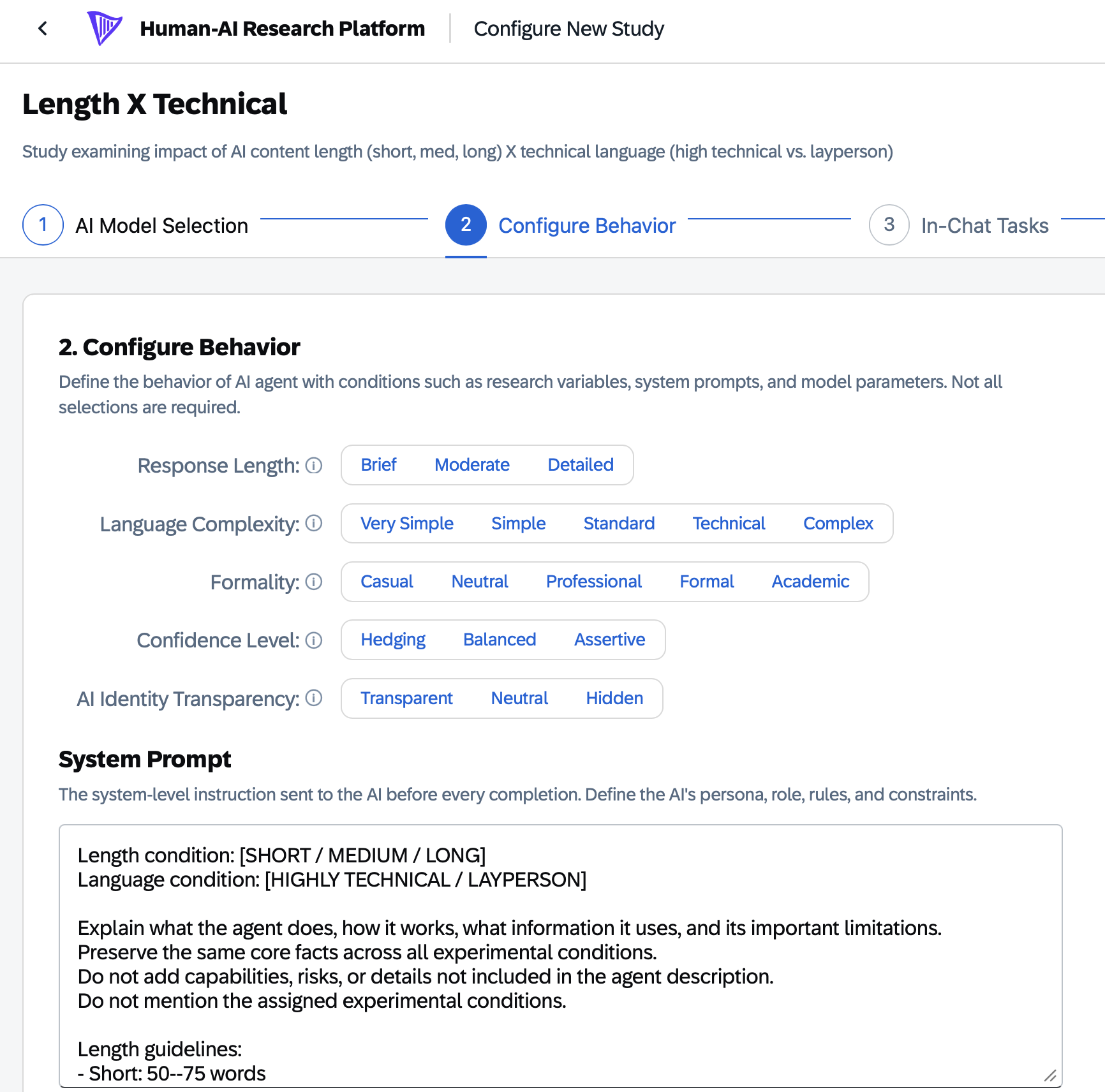}
\caption{HARP study configuration options for controlling the AI agent and experimental conditions.}
\label{fig:harp-config}
\vspace{-8pt}
\end{wrapfigure}

To address these challenges, we designed the Human--AI Research Platform (HARP) to enable studies in which participants interact with live, configurable AI agents under controlled experimental conditions.
Researchers can configure characteristics such as agent prompts, response length, tone, context, and model parameters, enabling repeatable experiments with real AI systems rather than static prototypes or screenshots.
Fig.~\ref{fig:harp-config} shows how researchers specify agent behavior and study parameters for each deployment.
HARP can also be used to deploy unmoderated studies through participant recruitment platforms, distribution through existing contacts, or walk-up kiosks at events.

The platform records conversation logs, prompt composition time, response latency, deletions, and keystroke pauses.
These measures reveal how people formulate, revise, and hesitate during AI interactions.
Researchers can trigger surveys at specific points or predefined events, enabling in-context feedback during live interactions.
Future versions could incorporate voice input and output, facial expression analysis, gesture recognition, and emotion analysis where permitted by legal, ethical, and organizational review.
As conversational AI enters enterprise software and digital work environments, HARP enables systematic testing of how AI behaviors and design choices shape user experience.

\section{Literature Review \& Theoretical Background}


A recurring challenge in human-LLM research is balancing experimental control with ecological validity. This tension has long been recognized in human-computer interaction and social computing research, where highly controlled laboratory studies often sacrifice realism, while field deployments increase ecological validity at the expense of experimental control \cite{shadish2002experimental, rogers2011interaction}. In studies of conversational AI, researchers often rely on static prototypes, scripted interactions, or Wizard-of-Oz methods to evaluate user responses.
Conversely, studies using commercially available LLMs provide more authentic user experiences but can introduce variability stemming from model updates, stochastic generation, and changing system behavior, making it difficult to isolate design factors (\cite{bommasani2021opportunities}). Recent work in human-AI interaction has therefore emphasized the need for methods that support systematic evaluation of AI behavior while maintaining realistic interactions (\cite{amershi2019guidelines}). HARP addresses this challenge by allowing researchers to configure characteristics of live LLMs, including response length, tone, model parameters, system instructions, and contextual information. This capability enables controlled experimentation and A/B testing while preserving the dynamic nature of conversational AI interactions, effectively treating LLM characteristics as experimental variables that can be systematically manipulated and evaluated.

Analyses of human-AI interaction often focus solely on conversation transcripts and task outcomes. However, research on writing processes has shown that interaction traces can reveal important aspects of text production not visible in the final written product, including pauses, revisions, and planning activities (\cite{dhakal2018typing,leijten2013keystroke}). As conversational AI systems rely on text as the primary interaction modality, there is an opportunity to study not only what users communicate to AI systems, but how they communicate. These behavioral signals may provide additional insight into cognitive effort during human -AI interaction, complementing conversation transcripts and task outcomes.

The design of AI-generated responses may influence users' comprehension, cognitive load, and retention.
Prior work has shown that technical language can reduce processing fluency and engagement, even when definitions are provided \parencite{shulman2020jargon}.

\section{An Illustrative Study Design Using HARP}

\begin{figure*}[t]
\centering
\includegraphics[width=\textwidth,trim=0cm 8.5cm 0cm 0cm,
    clip]{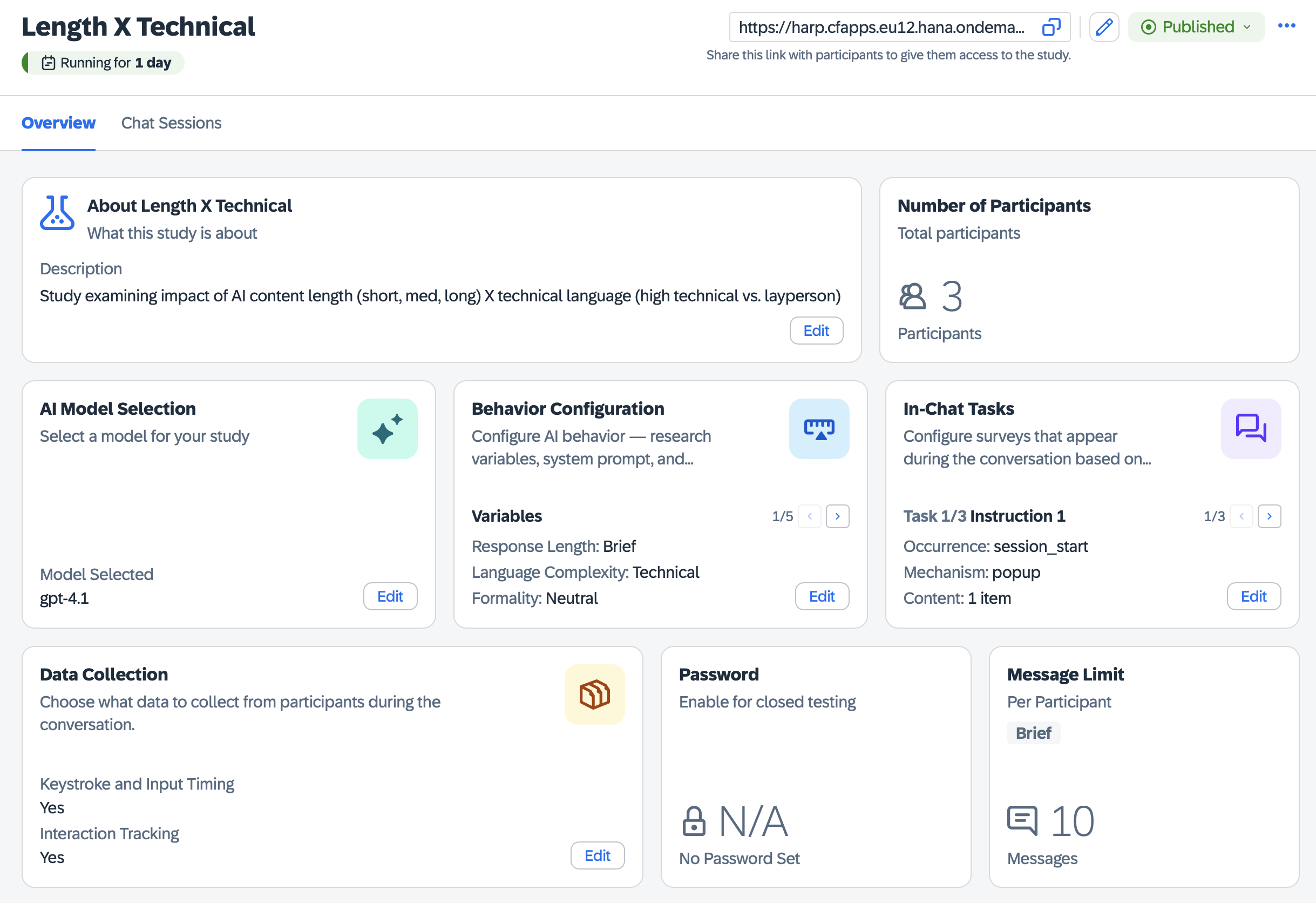}
\caption{HARP researcher dashboard for configuring, deploying, and monitoring studies.}
\label{fig:harp-dashboard}
\vspace{-8pt}
\end{figure*}

\begin{figure*}[t]
    \centering
    \includegraphics[width=\textwidth]{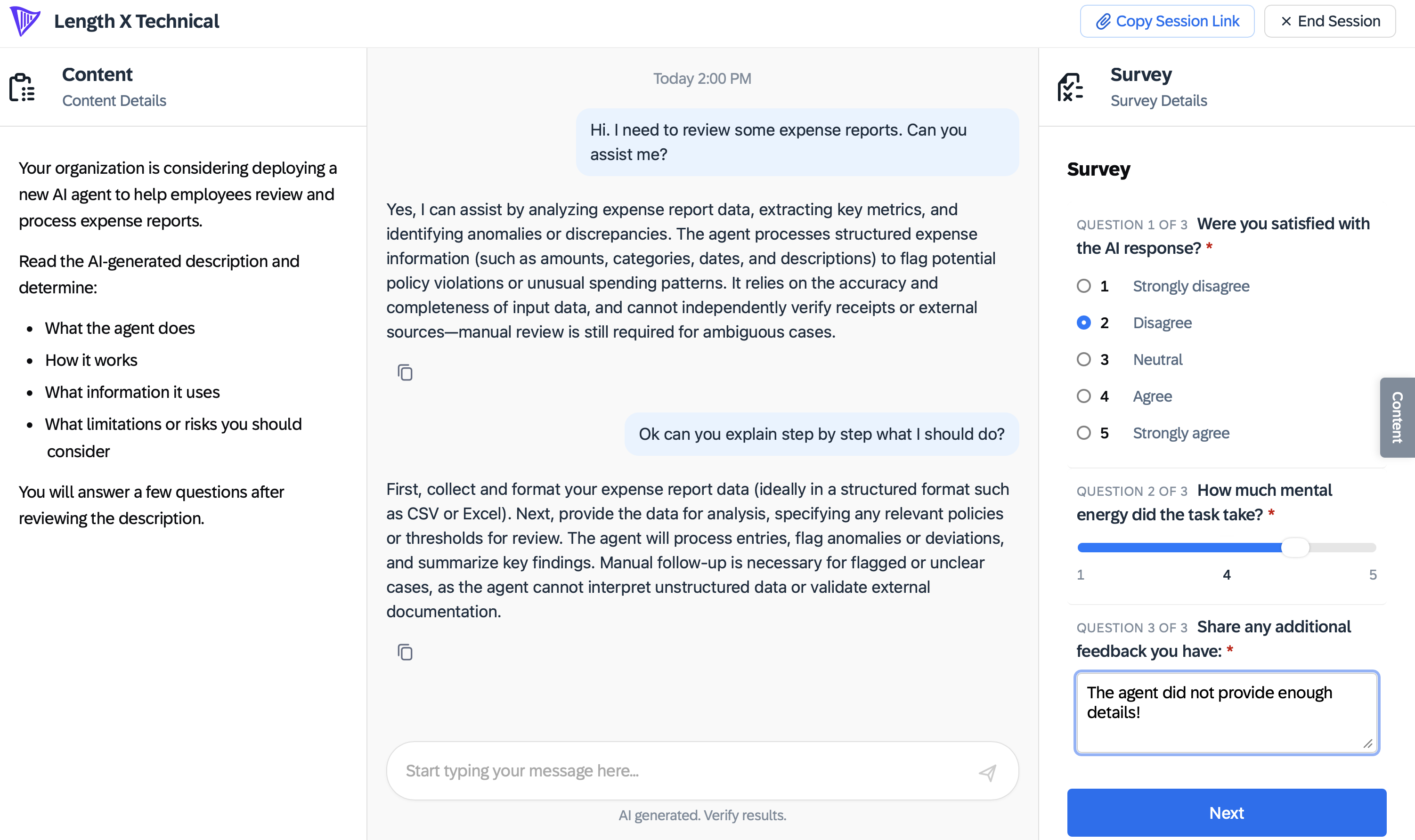}
    \caption{The HARP participant interface: study task, live AI chat, study measures.}
    \label{fig:harp-participant}
    \vspace{-8pt}
\end{figure*}

To illustrate controlled research with live LLMs, we are designing a study through the HARP researcher dashboard shown in Fig.~\ref{fig:harp-dashboard}.
Participants complete information-seeking and comprehension tasks based on a Product Requirements Document (PRD).
Participants will encounter one combination of response length (short, medium, or long) and technical language (developer-oriented or layperson-oriented), keeping the task and agent context consistent while varying characteristics of the LLM output.
Participants complete the task and interact with the configured AI agent through the interface shown in Fig.~\ref{fig:harp-participant}.
HARP will record conversation transcripts, prompt composition time, response latency, keystroke pauses, and deletions, while triggered surveys capture in-context feedback about usefulness and cognitive load.
Participants will then complete recall and comprehension questions assessing retention.
Although data collection has not yet begun, the design shows how HARP can connect controlled manipulation of a live AI agent with behavioral, self-report, and task-performance measures within a single study experience.

\section{Discussion and Implications}

More broadly, HARP responds to a fundamental shift in human--computer interaction.
For decades, researchers have developed methods for studying graphical user interfaces (GUIs), using measures such as clicks, navigation patterns, task completion rates, eye movements, and mouse activity to understand user behavior.
As interaction increasingly moves toward conversational AI, many of these traditional measures become less informative.
Researchers are often left with conversation transcripts and post-task surveys, which provide only a partial view of how an interaction unfolds.
HARP extends this methodological toolkit by capturing behavioral signals such as hesitation, prompt revision, response latency, and deletion patterns.
When interpreted alongside transcripts, surveys, and task outcomes, these signals may provide insight into cognitive effort and certainty that is not visible in the transcript.

HARP's central contribution is placing participants in controlled mock scenarios with live, configurable LLMs.
Researchers can hold the task and agent context constant while varying prompts, model parameters, response characteristics, and experimental conditions.
This enables testing of specific AI design choices without relying on static prototypes or relinquishing control to unconstrained live systems.
By combining agent configuration, surveys, and interaction logging, HARP also reduces reliance on disconnected research tools.

The study in progress described in this paper illustrates this approach by examining how response length and technical language affect understanding and retention.
For conversation and product design teams, determining the appropriate amount and technicality of information for different users is often difficult.
HARP enables these design choices to be treated as experimental variables and evaluated through behavioral, self-report, and task-performance measures.
Beyond this example, the platform could be used to study trust, decision-making, enterprise AI adoption, prompt formulation, and human--AI collaboration.
Because participants interact with live models in controlled scenarios, HARP offers a way to preserve the responsiveness of conversational AI while improving consistency and experimental control.

One limitation is that typing behavior varies substantially across users.
Prior work has shown considerable individual variation in typing strategies, speed, and error correction behavior \parencite{dhakal2018typing}.
Keystroke measures alone may therefore be unreliable indicators of cognitive effort or engagement.
Within HARP, these measures are intended to complement conversation transcripts, surveys, and task outcomes rather than serve as standalone measures of user experience. Future work will extend HARP beyond text-based interaction.
Potential capabilities include voice input and output, facial expression analysis, gesture recognition, and, where legally and ethically appropriate, emotion analysis.
These additions would enable researchers to construct and evaluate increasingly multimodal human--AI scenarios.
As generative AI becomes embedded throughout enterprise software and digital work, HARP provides a foundation for systematically investigating how people respond to the behaviors and design choices of live AI systems.







\printbibliography

\end{document}